# Cities of Indexes:

Articulating Personal Models of Urban Preference with Geotagged Data

ALVAREZ-MARIN, Diana and SALDAÑA OCHOA, Karla

[a]ETH Zurich, Institute of Technology in Architecture, Chair Digital Architectonics
alvarez@arch.ethz.ch, saldana@arch.ethz.ch

ABSTRACT

How to assess the potential of liking a city or a neighborhood before ever having been there? The concept of urban quality has until now pertained to global city ranking, where cities are evaluated under a grid of given parameters, or either to empirical and sociological approaches, often constrained by the amount of available information.

Using state-of-the-art machine learning techniques and thousands of geotagged satellite and perspective images from diverse urban cultures, this research characterizes personal preference in urban spaces and predicts a spectrum of unknown likeable places for a specific observer. Unlike most urban perception studies, our intention is not by any means to provide an objective measure of urban quality, but rather to portray personal views of the city or "Cities of Indexes."

KEYWORDS

Machine Intelligence, Human Intelligence, Indexes, Big Data, Urban Perception

1. INTRODUCTION: QUALIFYING, LEARNING, ARTICULATING

> *"There is a certain standard of grace and beauty which consists of a particular relation between our nature, such as it is, weak or strong, and the thing which pleases us."*
>
> (Pascal, 1657)

As people move through the city, they produce, collect, share, store, leave traces of their daily activities. In cities, a tremendous amount of data is being produced either intentionally, via blogs, social media, discussion forums, news, messages, or unintentionally via online and offline interactions. With the development of social media and virtual communities, the perception of a city is presented as a curated experience or "virtual echo chambers," where enclaves of like-minded people consume information in strikingly similar ways (Bessi 2016). Alternative approaches to profiling through "virtual echo chambers" predict preferences by learning from user's interaction with the web. In 2018, Google Maps launched a newsfeed-like recommendations system based on "matches," which will suggest how likely a user is to enjoy a place. A machine learning algorithm generates a probability value based on the business information, a user's personal food and drink preferences, places previously visited, and past ratings. While this approach predicts "matches", it does so by accessing user's data without awareness of the former. This research aims to shift the perspective from the one of a passive and unaware user to the one of an active persona, having access to data before any specific profiling.

In parallel, web mapping services such as OpenStreetMap, Google Maps, Google Street View, and Bing offer a new set of data about urban spaces, presenting a top-down, impartial representation of the world. Data of this nature has been used in applications for computer vision in architecture, urban planning, urban economics and sociology. To name a few: the identification of distinctive visual elements from Street View data (Doersch et al., 2012), the recognition of architectural elements of specific historical periods (Lee et al., 2015), the relationship among visual patterns of satellite images as well as streetscapes and commercial activeness (Wang et al., 2018), the identification of urban patterns at a large scale using Convolutional Networks and satellite images (Albert et al., 2017), and the development of regression models based on Street View images to predict socioeconomic indicators (Arietta et al., 2014 and Glaeser et al. 2015).

This research aims at bridging the gap between these two different narrations of the city. The first one, produced by citizens from a multiplicity of views and often referred to as "soft data" (Rouvroy, 2016), and the second one, appearing as an entirely objective representation of the physical aspects of the city. Therefore, how to project individual urban perceptions on top of The Generic City[1]? Which would be a citizen's favorite neighborhood, even before having ever been to a specific city?

In the 1960s, Kevin Lynch addressed the relationship between city form and what it meant to its inhabitants on his formative work The Image of the City. Here, he explored the impressions of

Boston, Jersey City and Los Angeles inhabitants about their neighborhoods, concluding that people formed mental maps of their surroundings consisting of five essential elements: paths, edges, districts, nodes, and landmarks (Lynch et al., 1962). For Lynch, this image of the city has three components: identity (the recognition of urban elements as separate entities), structure (the relation of urban elements to other objects and the observer), and meaning (its practical and emotional value to the observer).

Even if addressing a similar question, this research rather than trying to distinguish specific analytical elements, connected through relationship structures, considers that digital technologies offer a different kind of accessibility to the understanding of urban experience, which is outside of analysis. Unlike Lynch, this research does not attempt to explain urban perception analytically but probabilistically, this means by recognizing patterns of preference that can allow for prediction but without explicitly trying to explain why such patterns exist.

There are two main stages in the process: First, a generic stage, where urban patterns are detected from a dataset of 50,000 satellite images and 32,000 street view images using state-of-the-art machine learning algorithms. Secondly, a specific stage, where a fictional user is invited to label a generic dataset of street view images to train a Convolutional Neural Net (CNN)[2] classification model, from which individual maps of his preferences can be established.

In parallel to the two stages of this trip, the joint work of Human Intelligence and Machine Intelligence is essential: no Human Intelligence is able to scan and qualify such an amount of data completely nor coherently, yet its involvement is essential in the subjective process of articulating meaning. In contrast, Machine Intelligence under the form of image classification, feature extraction, and dimensionality reduction algorithms perform the task of fast big data computing. Ultimately, the application is at the interplay between both kinds of intelligence.

Rather than an objective collective view of attractiveness in cities that is exposed in state of the art literature, this application will produce qualitative maps presenting a personal disposition of urban preferences that can only be verified by the user herself.

### 1.1. Odysseus

Odysseus, a fictional user, is introduced in this paper as a traveler that will express his personal preference regarding some urban places. The interaction of Odysseus with the data based application will embody the conceptual interaction between Human Intelligence and Machine Intelligence. Ultimately, thanks to the predictions obtained from the application, Odysseus is offered recommendations of places he may like in multiple cities of his choice.

## 2. CONTEXT: THE PLENTY OF DATA

Mobile Computing and Urban Data Streams
Physical presence and temporal linear sequences are no longer exclusive conditions to urban experience. With the digitalization of urban life, a large part of the planet has become accessible within a few digitations. This phenomenon can be perceived in a two-fold manner.

Citizens are able to write narrations of the city and share them with other people through social media and mobile devices that reveal a new landscape of attention. The identification and orientation to the objective urban elements that Lynch introduced decentralize in a multiplicity of views that would make the preservation of a fixed frame of reference impossible. Moreover, while for Lynch, the visibility of landmarks was crucial in the construction of this common frame of reference, today with mobile computing and social platforms exchanging "soft data", the role of the citizen becomes primary in the constitution of multiple views of the city. Simultaneously, the city can be represented following established mapping techniques, as tremendous amounts of geo-referenced data are readily available through web mapping services. How to conciliate both visions of the city?

### 2.1. Indexical Characterization

An index is a pointer that refers to a specific object. Such an object can be described in a multiplicity of manners (that might be properties, locations, qualities, etc.) as of recently enlarged with the digitization urban spaces. Therefore, in the scope of this research, the concept of indexicality is considered as the ability to characterize anything, by pointing, addressing, and connecting objects, yet without trying to exhaust all of its possible indexes. With this in mind, this research does not operate on the level of things or places in themselves but rather among the indexes that characterize them.

To characterize a city or a space is a consistent yet never complete gesture. An urban space can be characterized n-dimensionally by its location, image information, ratings and the like. However, the descriptions that one observer might make are entirely different from the ones captured by a different observer. It is the citizen who fixes, chooses and defines answers. With this in mind, this experiment aims at opening up a discussion towards a more vibrant and personalized experience of global urbanity rather than a unified urban perception.

## 3. METHODOLOGY: ACCESSING THE PLENTY OF DATA

### 3.1. Taking A Subset from The Plenty of Data

Google Maps satellite images are fundamentally available for most of locations on the planet. To take a subset out of this plenty is not a neutral act. In order to follow the personalized approach explored in this experiment, the fictional traveler Odysseus will define a subset by selecting 20 cities from five different continents, provided each with a fair amount of accessible Google street view images (fig. 1). This selection is completely arbitrary as it pertains to Odysseus' personal interest and any variation in this subset will unfold into entirely different answers. This initial subset covers the following cities:

*Mexico City, Bogotá, Buenos Aires, New York, Los Angeles, London, Zürich, Paris, Moscow, Tokyo, Delhi, Taipei City, Istanbul, Bangkok, Kampala, Jerusalem, Johannesburg, Kumasi, Sydney,*

*and Melbourne.*

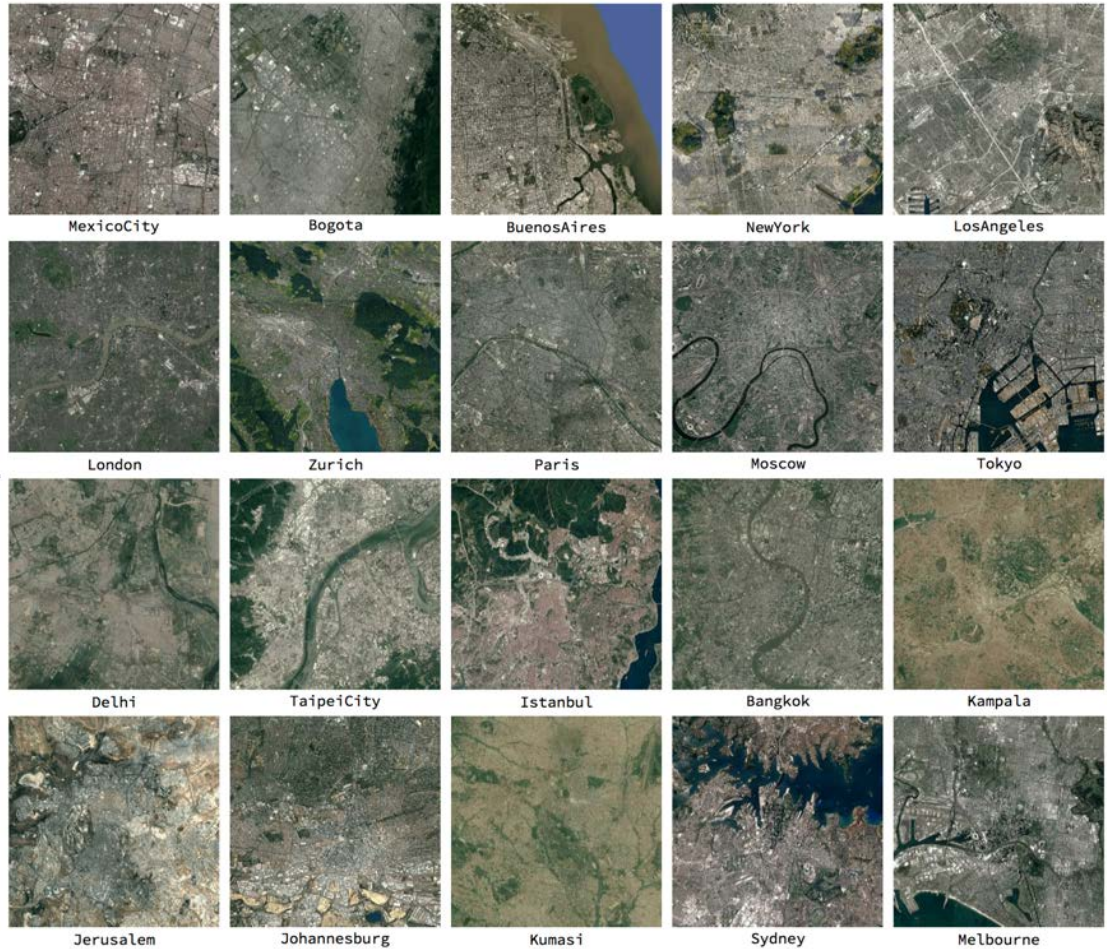

Figure. 1. Satellite images of 20 cities from five different continents, selected by Odysseus.

### 3.2. Creating A Unit of Measurement

Each one of these cities is partitioned into places, by defining a comfortable radius from which a an observer placed at the center of its circumference can visually perceive the totality of a place. Therefore, each city covering an area of 10x10 kilometers is subdivided into a regular grid of 2,500 places of equivalent areas of 200x200m with no overlapping, crawled at a zoom 18 in the WGS 84 projection with the Google Maps API.

Odysseus dataset sums up now to 50,000 satellite images, by combining the 2,500 places of each of the 20 cities. Using these 50,000 geolocations, a partial set of 32,000 corresponding street view pictures are collected. While satellite pictures offer an objective top view, street view pictures offer a perspective eye-level view, more alike to the perception of space that an observer may have while being physically in a place. Street view pictures are used as training data, by being labeled with a rating of preference provided by Odysseus. This process is explained in detail in the chapter HUMAN INTELLIGENCE: ODYSSEUS RATING OF URBAN SPACES.

Scale correction is necessary so that these 50,000 satellite images are comparable (fig. 3), presenting the same scale and covered surface. All satellite images are corrected regarding latitude deformation (fig. 2) using the following formula (1), where 6378137 refers to the radius of the Earth in meters.

$$groundresolution = \frac{Cos\left(latitude \cdot \frac{pi}{180}\right) \cdot earthcircumference}{mapwidth}$$

*ground resolution (in meters by pixel)*
*earth Circumference = $2 \cdot pi \cdot 6378137$*
*map width = $256 \cdot 2^{zoomlevel}$*

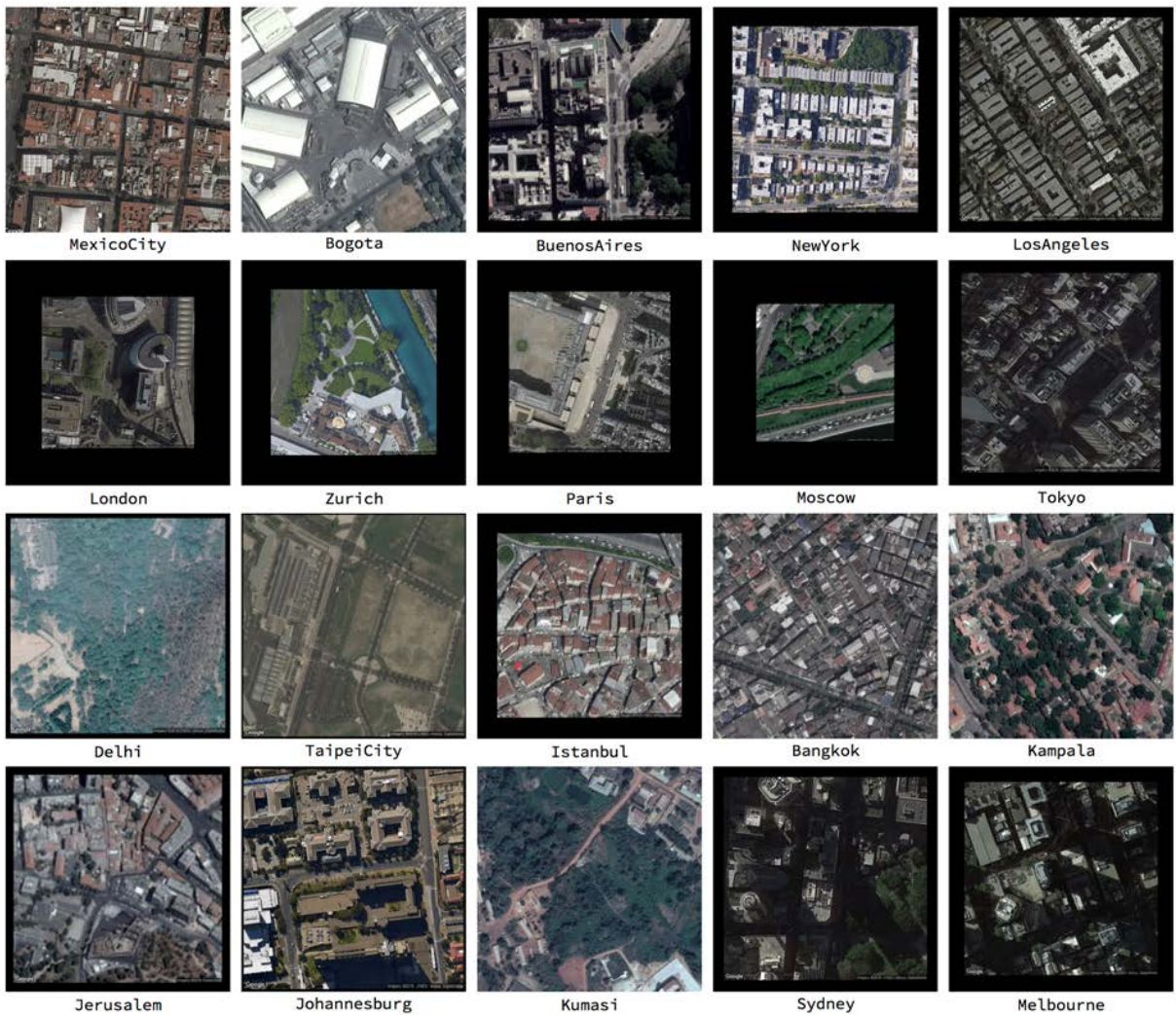

Figure. 2. These sample images before deformation correction show differences among covered surfaces while using the same scale (zoom 18 for Google Maps). As latitudes get further away from the Equator, the covered surface shrink.

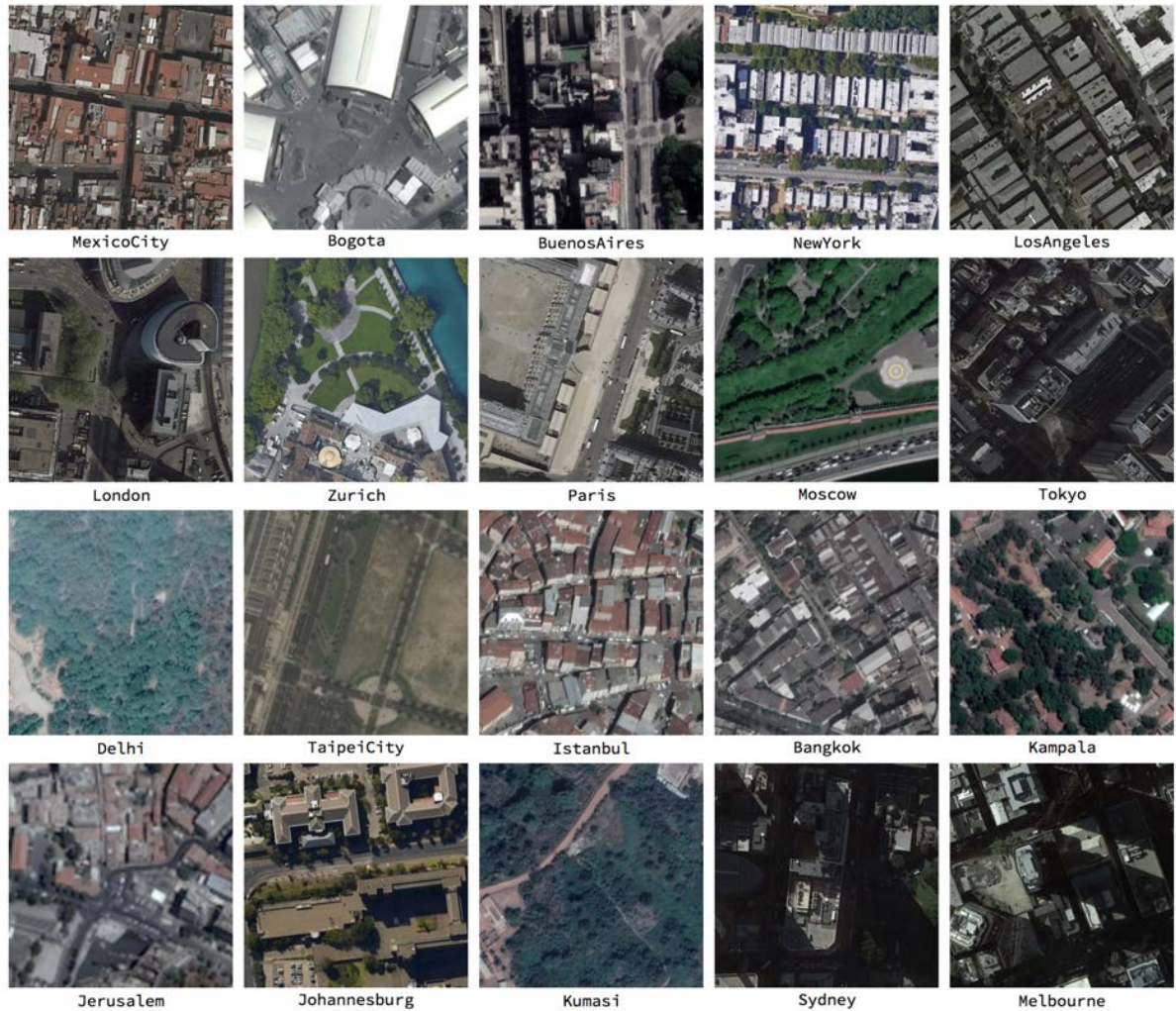
Figure. 3. These sample images after deformation correction have a larger zoom so that all latitudes, especially the ones which are more distant to the Equator, present the same covered surface.

The totality of this dataset, all 50,000 satellite images combined appears at first as a collection of raster images in a disorganized state, that represent physical objects, buildings, streets, trees or rivers; however the diversity of this content makes difficult the recognition of similarities. These images are transformed in numerical vectors able to encapsulate their features in terms that allow for comparison. This digitized symbolization, under the form of numerical vectors, no longer pertains to the field of representation, reachable through the senses, but can be computed thanks to Machine Intelligence.

As pointed out by Rouvroy, the "units" of Big Data are neither individuals nor objects, but data (2016). The units of Big Data emerge out of data itself in a self-referential process wherein the measurement is performed in inextricable relation to the whole subset which is being measured. For Barad, these measurements are performative agential practices, this means that they are a constitutive part of what is being measured and are inseparable from that which is observed. Measurements are world-making, she adds, matter and meaning do not preexist, but instead, are co-constituted via measurement intra-actions (2012).

The self-referential numerical encapsulation of each picture in relation to all the others is implemented by a process called feature extraction. The latter translates each image into an n-dimensional vector, where the value of n depends on the given initial dataset that can be latter on projected to a lower dimensional space (fig. 4). To perform the feature extraction, a feature extractor model (FE) is designed by cutting off the last layers of a trained VGG network (stack of convolutional layers, followed by fully connected layers) (Simonyan et al. 2008) and using the convolutional layers (feature extraction part).

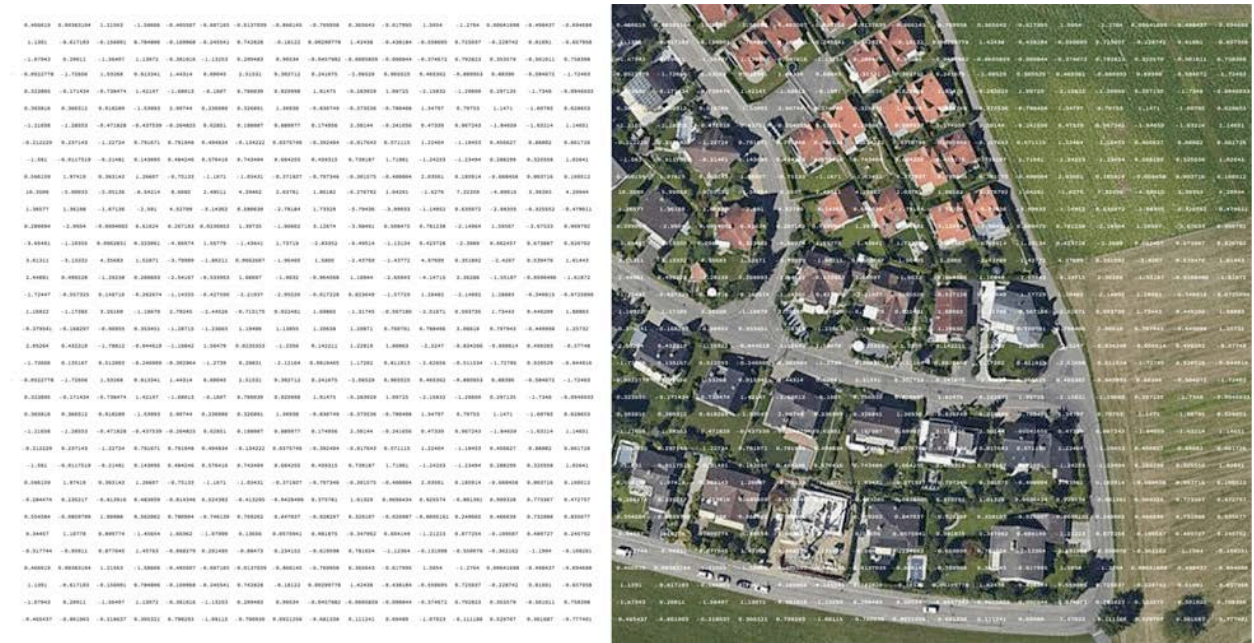

Figure. 4. Symbolic encapsulation of a satellite image into a n-dimensional numerical vector.

Pre-trained CNN models[3] are used as a benchmark to either improve an existing model, or test another model against it. The pre-trained model (ImageNet, VGGNet) used in this experiment is trained with a large number of labeled images used in other visual recognition tasks. The first trained layers of this model are fixed and they perform the task of feature detection. In this way, features of each image are extracted, considering a subsample of 20,000 pictures (about 40% of the subset), resized to 224 × 224 pixels in 3 RGB bands, to fit the architecture of the model. The comparison of satellite images is a more complex task than the one of photos of specific objects, such as cars, faces or cats, due to their greater structural diversity. Therefore, they are radically different from the images initially used to train our model (ILSCVRC2012, 2012), however, the latter is still able to learn from satellite images that were not included in its initial training data.[4]

Through this process, images are transformed into feature vectors of 4096 dimensions, encoding high-level concepts, such as green areas, suburban fabric, industrial spaces, and the like. These concepts are more complex and less specific than the usual physical land features integrated into a satellite image, such as trees, buildings or roads. The vector representation produced by this model encapsulates semantic meaning in a "meaningless" way, being also independent of specific pixels' values in the input image (fig. 4).

The previous process shows the interplay between the architecture of the model and data. On a more abstract level, it supports the idea that the act of measurement is a constitutive part of what is being measured, as the model "learns" from the data it is being fed with.

### 3.3. Generic Recognition of Urban Features

The initial dataset of pictures is now a list of 50,000 vectors of 4096 dimensions each. Such a high dimensionality needs to be reduced so that a Human Intelligence can make an interpretation out of it. After operating and computing on a high dimensional level, vectors can be reduced to a two-dimensional space using the dimensionality reduction algorithm t-SNE (van der Maaten, 2008). This is how n-dimensional vectors, meaningless to humans, can be transferred again to a lower dimension, where a stimulating arrangement of spatialities can be opened up for personal interpretation.

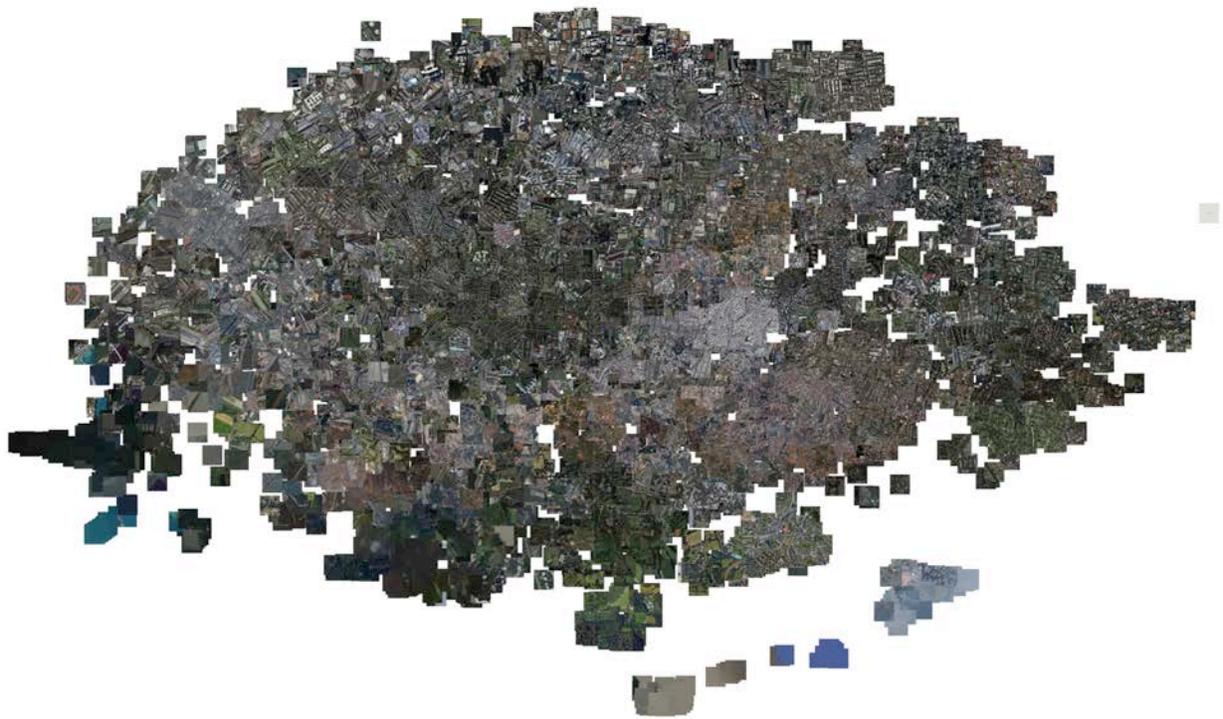

Figure. 5. A cloudy arrangement of pieces of cities after dimensionality reduction to two dimensions, projected on a Cartesian plane.

t-SNE tends to centralize the position of the data. To adequately explore relationships in the data, this cloudy arrangement (fig. 5) can be unfolded into a continuous two-dimensional map or landscape of similarities by using the unsupervised machine learning algorithm Self Organizing Map (SOM)[5] (Kohonen, 1982), where each image positions itself based on its similarity to its neighboring images (fig. 6). The SOM is initialized with an 80x80 cells matrix and fed with an input of 50,000 two-dimensional vectors, obtained in the previous step. The training procedure involves one million epochs until the response layer evolves into a stable configuration.

The resulting spectrum spans over 6,400 cells or Best Matching Units (BMU's) that can be considered as characters inside an alphabet, where each character impersonates a certain spatial quality without needing to refer to a specific position in space.

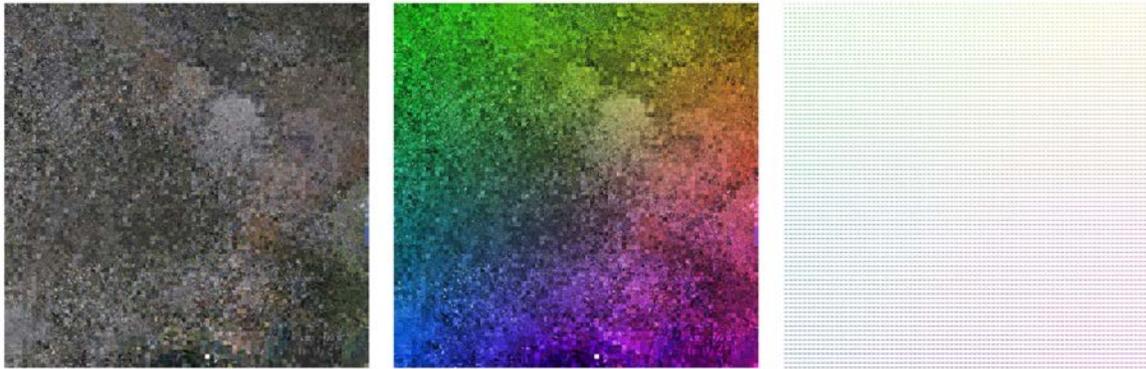

Figure. 6.  Generic SOM: Clustering of spatialities, spectrum of the BMU's weight values and alphabet representing each one of the BMU's with an assigned character.

The implementation of Machine Learning Algorithms in remote sensing and land use classification is still recent, and state-of-the-art approaches tend to define a specific number of classes to represent a particular city. Yang et. al (2010) suggests using the UC Merced land use, a dataset of 2,100 images spanning over twenty-one classes. In 2001, the European Union defined an Urban Atlas (2011) containing twenty classes of urban land use, this atlas is defined as a standardized dataset for land use classification. Albert et al. (Albert et al. 2017) uses ten classes from this Atlas, in order to classify land use in ten European cities. TerraPattern [2016], Descartes Lab [2016], and DeepOSM [2016] provide search tools for satellite images based on predefined land use classes. In their approach, satellite images from Google Maps are being paired with truth labels over a large number of specific classes obtained using the Open Street Map API. The classes they employ are particular, e.g., baseball diamonds, churches, or roundabouts and are posteriorly applied to the whole dataset using a supervised convolutional architecture for classification.

Unlike state of the art, for this experiment specific labelling is avoided, so that the whole spectral alphabet of 6,400 characters can be preserved. With this in mind, a grouping of several characters could correspond to words that can describe a city as text. For this purpose, rather than labelling, a color spectrum or non-semantic labels can be assigned to such characters. Characters can be traced back to a position in space if desired, so that recovering their spatial order they can be seen as a specific texture of colored pixels or spatialities constituting a city fabric (fig. 7).

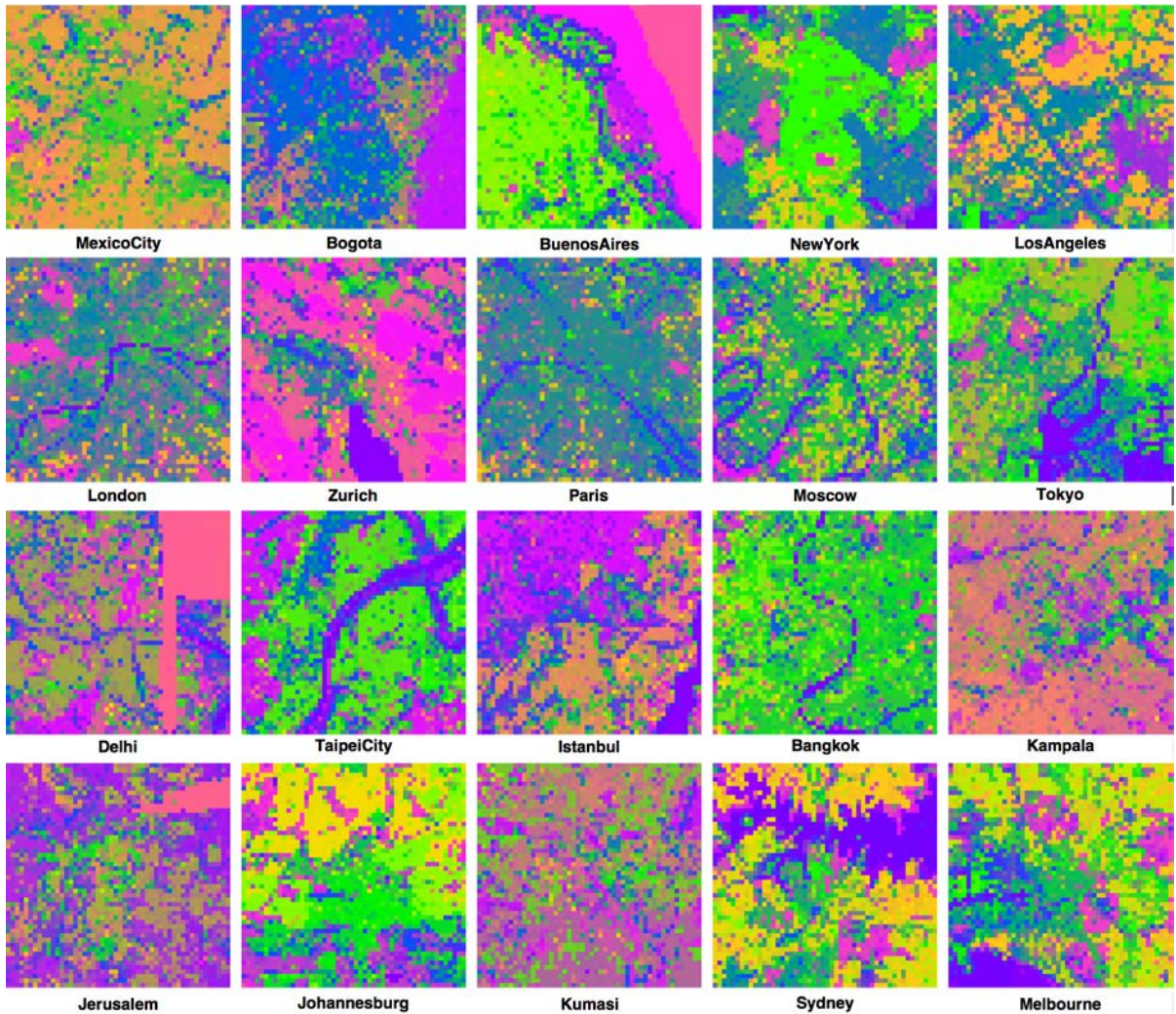

Figure. 7. Generic Pixel Maps of cities produced after projecting characters on their corresponding geo-location show predominant spatial types.

In order to find urban patterns and similarities among the cities in the subset, once again the image features of these pixel maps are extracted and visualized in a two-dimensional space using t-SNE (fig. 8). Each city finds its position in this space based on its similarity to its neighboring cities, comprehending both urban pattern and structure. In this manner, their spatial typicalities and exceptions can be highlighted. These 20 cities, in the beginning in a disordered condition can now cluster based on their spatial and structural similarities.

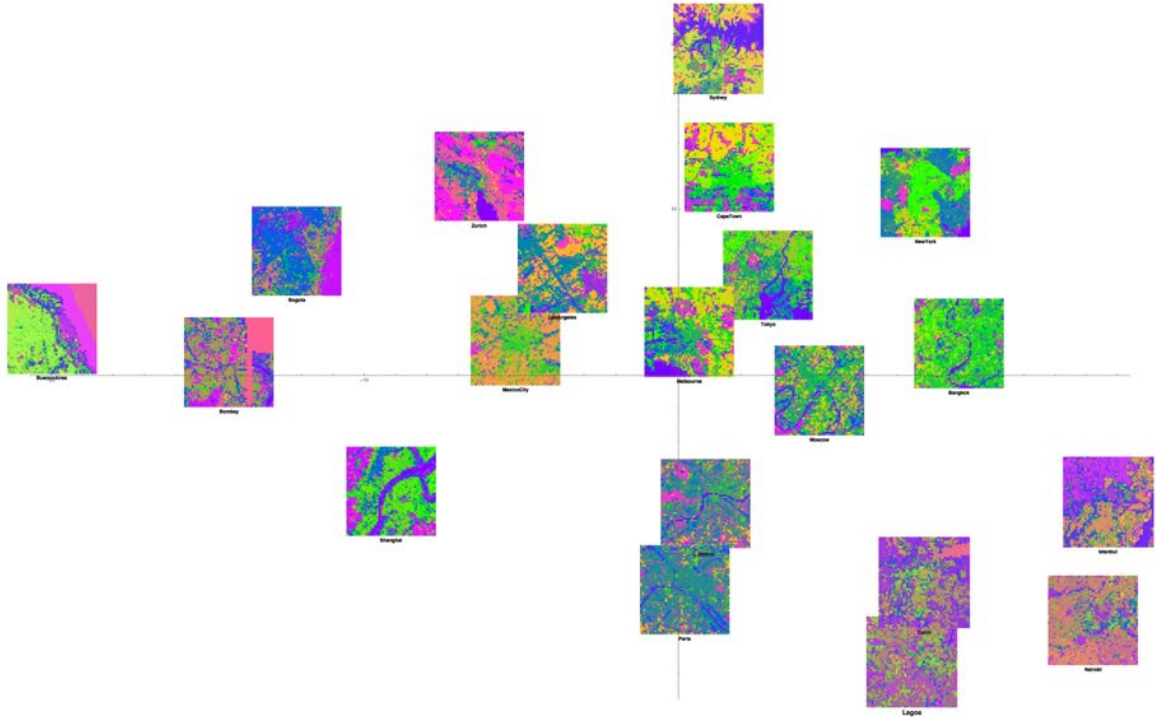

Figure.8. T-SNE clustering of 20 cities based on spatial and structural similarity.

In the image above the formation of clusters such as Mexico City and Los Angeles, London and Paris, Cairo and Lagos, Istanbul and Nairobi, can be noticed. Using the obtained alphabet with its 6,400 characters (spatialities) and projecting each of them on its corresponding geolocation (structure), similarities and differences were found, without prescribing any urban land use, density, nor socio-economic factors. Subsequently, Human Intelligence will be integrated by inviting Odysseus to rate street views based on personal preference.

### 3.4. Human Intelligence: Odysseus Rating of Urban Spaces.

Until now these places have been considered only from their top view. However, perspective images provide a more sensorial perception of space that can be judged by personal preference. Odysseus rates the Google street view images, while disregarding anecdotic conditions, such as weather, lighting, focus of the camera, etc; and instead concentrates on the intrinsic qualities of the space. To create a representative sample of images for Odysseus to choose from, the totality of street view images is grouped into clusters of similar features. As previously explained, features from images are extracted and fed to a dimensionality reduction algorithm (t-SNE), so that 32,000 vectors of two dimensions are produced. In the figure below, the approach seen in the previous chapter is repeated. The initial centralized t-SNE clustering can be fully unfolded in a two-dimensional SOM of 80x80, achieving 6,400 BMUs (fig. 9).

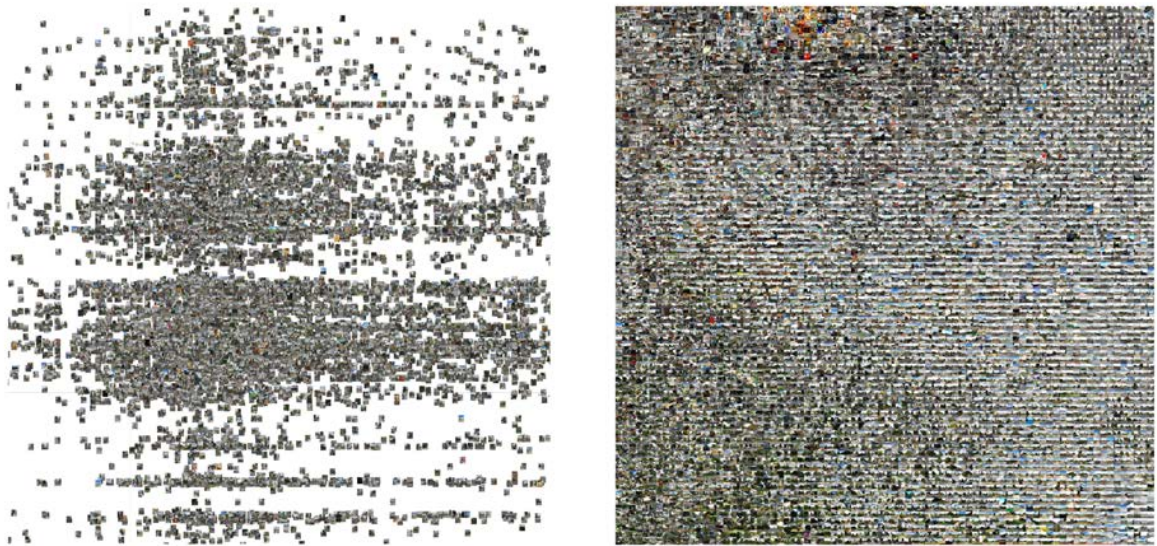

Figure. 9. Two-dimensional T-SNE and its unfolded projection with SOM.

The K-means (Hartigan, 1979) clustering algorithm is fed with the weight vectors of the 6,400 BMUs resulting from the trained SOM. A maximum number of clusters out of this map is obtained, resulting in 513 cases out of which each centroid becomes a representative element or a condensed way of conveying information (fig. 10). Vesanto et al. (2000) proved that this process performs better than a direct clustering of the totality of the dataset (32,000 street view images) and reduces computation time. With this approach, a compact, representative, and reliable training dataset of 513 images can be efficiently labelled by Odysseus without losing consistency.

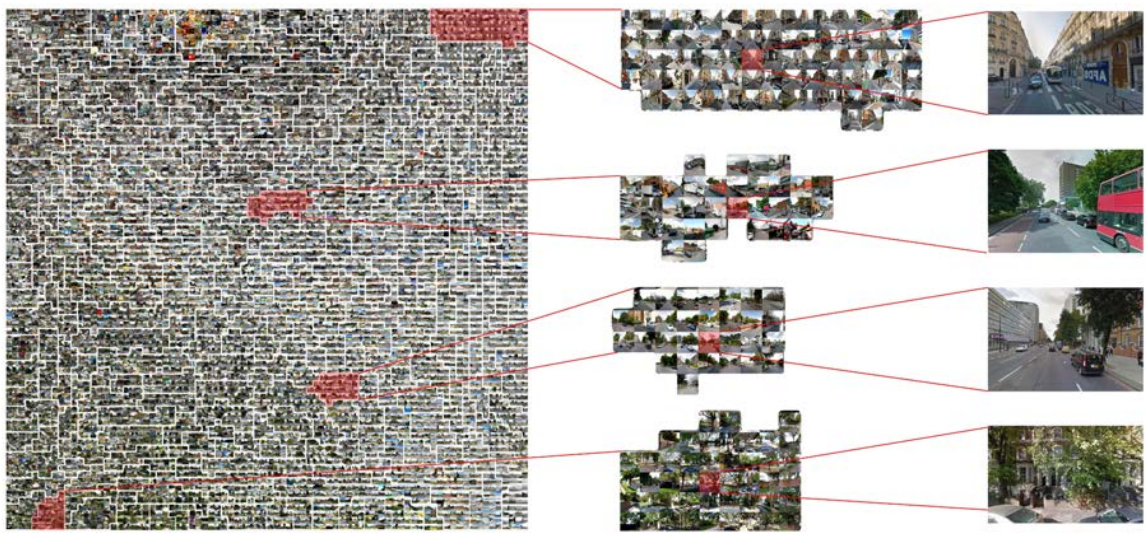

Figure. 10. K-means clustering algorithm on the SOM, sample clusters and their centroids.

By randomly iterating over the 513 centroid images, a survey of 1,500 pairs of images is created in order to be rated by Odysseus. Therefore, rather than obtaining a single numerical score for each picture, a relative comparison between pairs of images is performed, to obtain a persistent and consistent personal vote[6]. Each image is shown at least three times in comparison with another, allowing Odysseus to grade the pictures several times and in different pairings. Only places that have at least two or more favorable votes are considered as attractive.

### 3.5. Machine Intelligence: Image Classification

The conventional approach towards classification is to have a universal classifier or one model trained with data labelled by people, usually on crowdsourcing platforms such as Amazon, Mechanical Turk, Reddit, or academic projects surveys. Authors such as Quercia et al. create a dataset of votes from over 3.3K individuals and translate them into quantitative measures of public perception (2014). Dubey et al. introduce a crowdsourced dataset containing 110,988 images to produce public perception data at the global scale (2016). This approach is not recent, already in 1967, Peterson proposes a quantitative analysis of public perceptions of neighborhood visual appearance, having 140 individuals rate 23 pictures of urban scenes in the Chicago. These studies propose a global dataset that presupposes the existence of a common sense, which encapsulates the individuality of preference and perception. This research wants to invert from the approaches presented above by focusing on the personal preference of Odysseus rather than investigating a collective understanding of likeable and dislikeable places, as seen in state of the art approaches.

Rahwan from Scalable Cooperation, has shown in its AI-Powered Psychopath, that data can significantly influence the behavior of a machine learning algorithm (Norman 2018). The training data is as essential as the architecture of the model. Taking this into account, a new machine learning model is trained with places Odysseus likes and dislikes.

The dataset and the architecture of the model are prepared according to the following pipeline. The dataset is split in a 60–5–35 ratio for training, validation, and testing. To do so, popular data augmentation techniques are implemented, including random horizontal and vertical flipping, random shearing, random scaling, random rotations of the input images (Melo et al. 2017), having as a result 3,600 labelled images. Once again, the VggNet model is implemented, yet this time in a supervised manner for classification purposes, by preserving the architecture of the model until its 5th Pooling Layer. The annotated dataset is then used as training data in the classifier mentioned above. Three Fully-Connected (FC) layers are added posteriorly: the first with 1000 channels, the second with ten channels, and the third with two channels (one for each class) fig. 13. The final layer is a soft-max layer that outputs a probability distribution that sums up to one. All hidden layers are equipped with the non-linearity rectification ReLU (Krizhevsky et al., 2012).

As performed previously, the input images had 224 × 224 pixels in size with 3 RGB bands to fit the architecture of the model. The stochastic gradient descent is used to optimize the network loss function, starting with a learning rate of 0.1, and halving the rate every ten epochs. These networks are trained for at most 100 epochs, with 100 samples in each epoch, stopping the learning process when the accuracy on the validation set does not improve for more than ten epochs. The training process can be performed endlessly and every time the model is trained it gets rearticulated: given a certain input it is able to predict a unique classification, as a probability between the learnt classes: like (1) and dislike (0).

After this model is trained and new vectors are obtained, a classification of the 32,000 perspective images based on their ratings can be represented as a SOM of 6,400 BMUs, as seen in the previous chapter. Ultimately, a spectrum of places spanning from "likeable" to "dislikeable" is probabilistically predicted. It is worth stressing that Odysseus has never experienced most of these places physically. Every resulting spectrum is unique and comes as a result of the interaction between the model and the data.

## 4. THE APPLICATION: SUGGESTING NEW PLACES TO VISIT.

Satellite data is largely available, in opposition to the limited availability of perspective views of any place on the planet. Both satellite and street view images used in this experiment have geo-location as metadata. This common point or index, allows to transfer the label assigned to the 6,400 BMU's of the perspective images to their spatially corresponding satellite images, in order to create a new training dataset. This process is known as domain adaptation[7] or model portability, where the predicted labels from a model that is trained with a specific dataset can be transferred to a new dataset. This model is fed with the 43,600 remaining satellite images. As a result, 50,000 labelled images are obtained with a probability of being liked or disliked, by Odysseus. By adapting the domains, a personalized model to classify the world is articulated.

Together, these 20 cities create a learning base for those spaces Odysseus could potentially like or dislike. The model used to classify the satellite images has the same architecture as the one used to classify the perspective views. The stance of this research is not to look for reason or logic in personal preference but rather to predict probabilistically by learning from thousands of examples. This process of learning can reach a certain consistency in its predictions[8] but can hardly be considered as complete, since the amount of parameters that could possibly explain personal preference is inexhaustible.

Specific Prediction of Urban Patterns Based on Preference

> *"Words differently arranged have a different meaning, and meanings differently arranged have different effects."*
>
>  *(Pascal, 1657)*

The liking or disliking of a place is not entirely a binary choice but a probability inside a spectrum, as things can be liked or disliked a little or a lot more. Out of this classification, a characterization for each of the 50,000 points is obtained, from which some patterns may emerge. These probability patterns are topologically located based on their similarities, appearing ultimately as a spectrum of personal preferences.

The procedure seen in the chapter The Generic Constitution of Things, where characters of an alphabet are projected on space to produce pixel maps, is repeated. However, this time, both generic urban patterns and Odysseus' specific preferences are projected together. To achieve this, the 50,000 satellite images are characterized simultaneously in these two manners. Two linear SOMs of a one-dimensional grid topology of 10,000 cells are trained, produce what is known as "contextual numbers"[9]. These two lists of contextual numbers are transposed into a single list of two-dimensional vectors to be fed to a new SOM of 80x80 cells, producing a new alphabet of characters, that encapsulates this time, urban pattern and preference simultaneously (fig. 11).

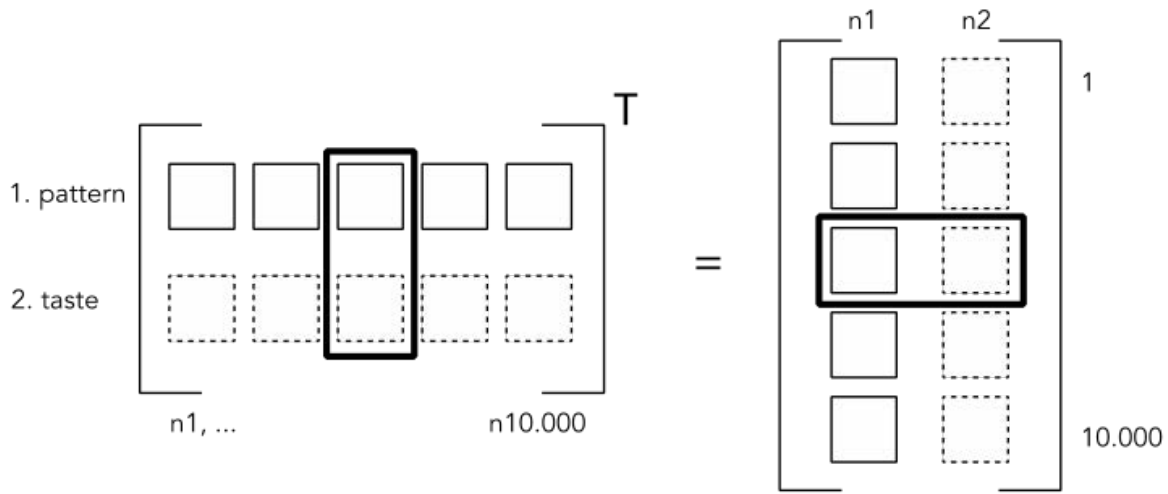

Figure 11. The two one-dimensional of contextual numbers are transposed into a two-dimensional matrix.

This time a linear spectrum spanning from warm to cold tones, allows a clearer rendering of the probability distribution between the "liking" and "disliking" of a place. Odysseus' patterns of preference are reaffirmed in a spectrum, where colder colors indicate preference and warmer ones point towards aversion (fig. 12).

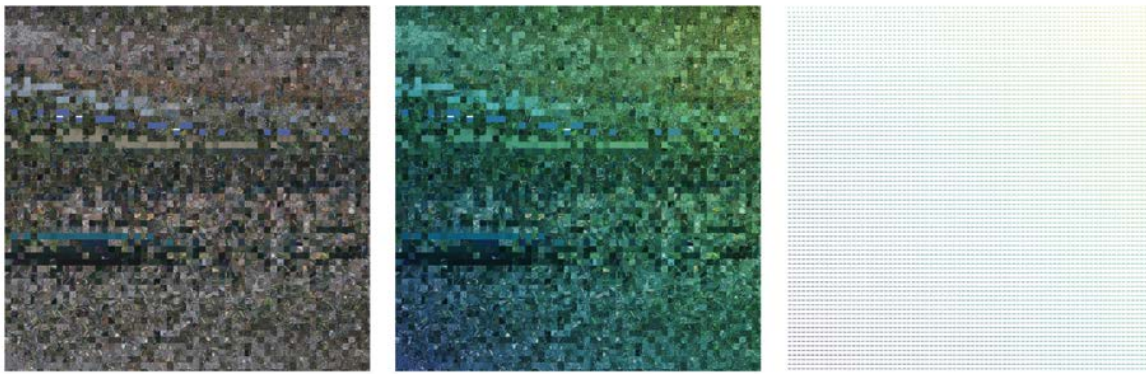

Figure.12. Specific SOM: Clustering of spatialities and preferences, spectrum of the BMU's weight values and alphabet representing each one of the BMU's with an assigned character.

By articulating Odysseus' preference and urban patterns together, a set of specific pixel maps is generated. These specific articulations of cities are not cities themselves but only Odysseus' personal interpretation (fig. 13).

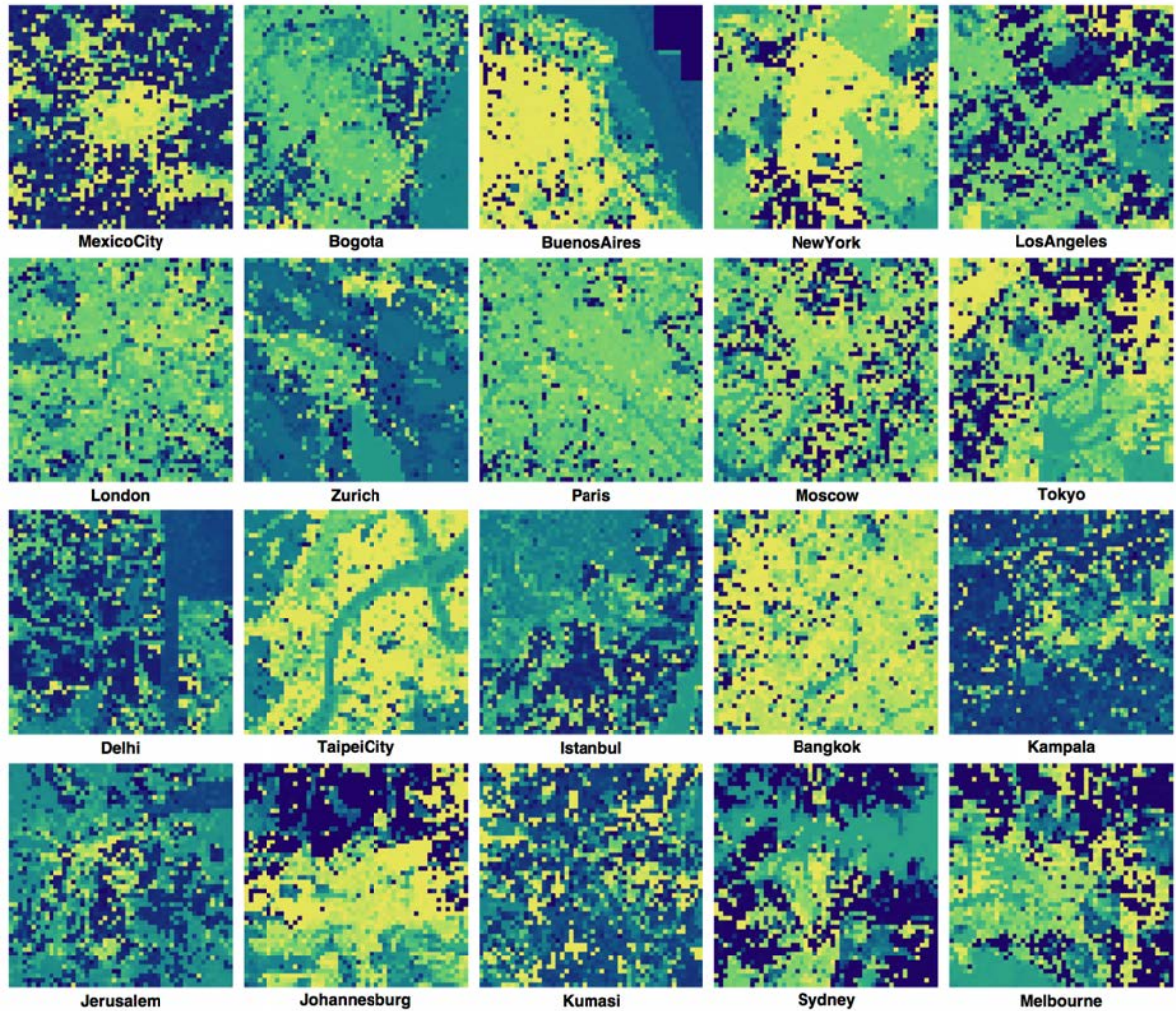

Figure. 13. Specific Pixel Maps of cities presenting a spectrum based on urban structure and personal preference.

Using the t-SNE algorithm, once again the resulting "specific pixel maps" are clustered so that cities liked for similar reasons come together; this means that they share simultaneously spatial structure and preference (fig. 14). In the figure below, several associations of urban patterns and preference emerge as the distribution of cold and warm colors varies in density.

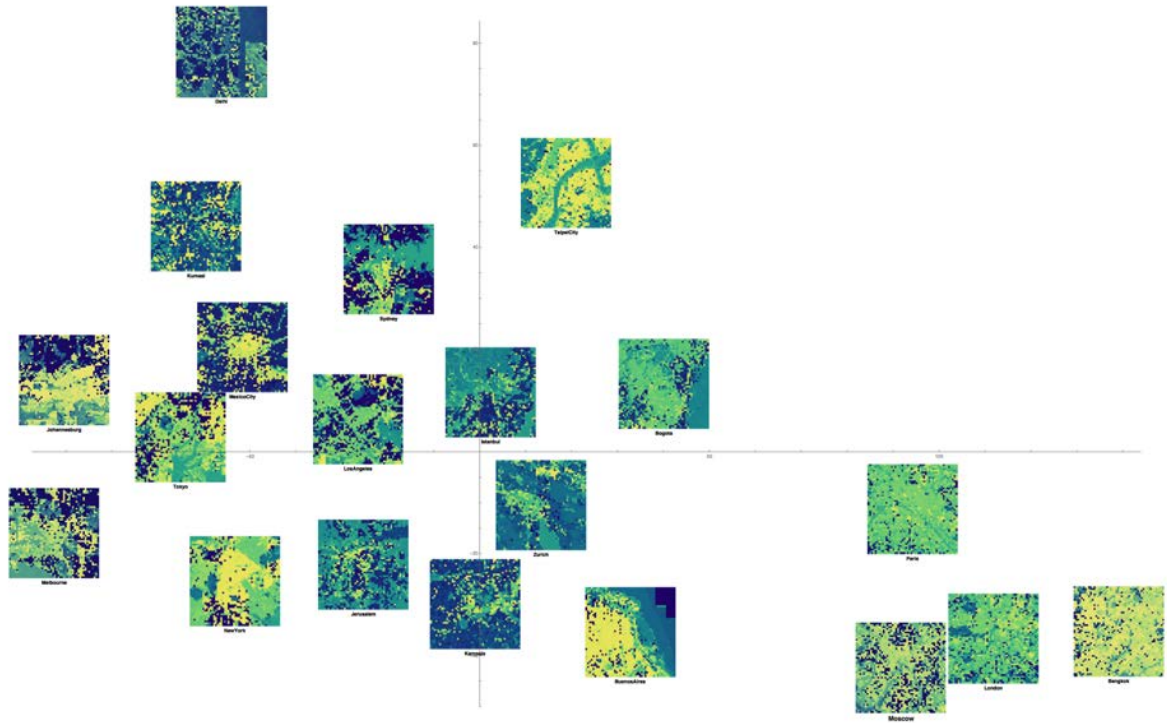

Figure. 14. T-SNE clustering of 20 cities showing spatial structure and Odysseus's affinities.

In cities like Moscow, Paris, and London, on the bottom right, the distribution of preferred spaces is scattered all over the city and appears punctually. These places correspond to public parks, boulevards, or planted streets, squares and waterfronts. At the other side of the spectrum, cities like Delhi and Kumasi, with predominant green areas such as large forests, show dense concentrations of likeable places with locally scattered unpleasant ones (fig. 14).

In the center left (fig. 14), appears a different typicality, where cities like Johannesburg, Tokyo, Mexico City, Melbourne, and New York present a clear division between large areas of likeable and unpleasant spaces. Bordering elements between these large areas could act as boundaries, and often correspond to the encounter among different urban patterns. It can be assumed that Odysseus likes hybrid rather than pure spaces, the encounter among elements of different kinds. For instance, he prefers boulevards and ramblas over forests and large parks, even if he also likes the latter. These cities are characterized by densely populated areas in contrast to large open green spaces.

Ultimately, cities such as Zürich, Istanbul, Jerusalem, and Kampala have a predominant spectrum of colder colors with some scarce dislikeable places which correspond to densely built areas, infrastructures and industrial areas (fig. 14). These cities contain a significant amount of spaces that Odysseus could potentially like. However, he might be able to find these anywhere on the planet yet in a different arrangement: as points, as boundaries and as areas.

## 5. DISCUSSION

The present experiment was not only an exploration of dealing with urban data streams but also a vivid example of communication among seemingly incomparable entities such as cities and places from different urban cultures. The experiment started with data collection by crawling two representations of the city, top (Google maps) and eye-level (Google street view) views. Satellite images were vectorized using a feature extractor, clustered using t-SNE and characterized as an alphabet of spatialities using SOM, as seen in the chapter Generic Recognition of Urban Features. Simultaneously, Odysseus performed a rating of urban places, as seen in the chapter Human Intelligence: Odysseus Rating of Urban Spaces. Based on these ratings, street view images were vectorized as probability values, as seen in the chapter Machine Intelligence: Image Classification. Until now these processes were performed in a parallel manner, so that two ways of perceiving a city were presented and used jointly to generate Odysseus' City of Indexes. In the following figure (fig 15.) a roadmap explaining the pipeline of the experiment is displayed

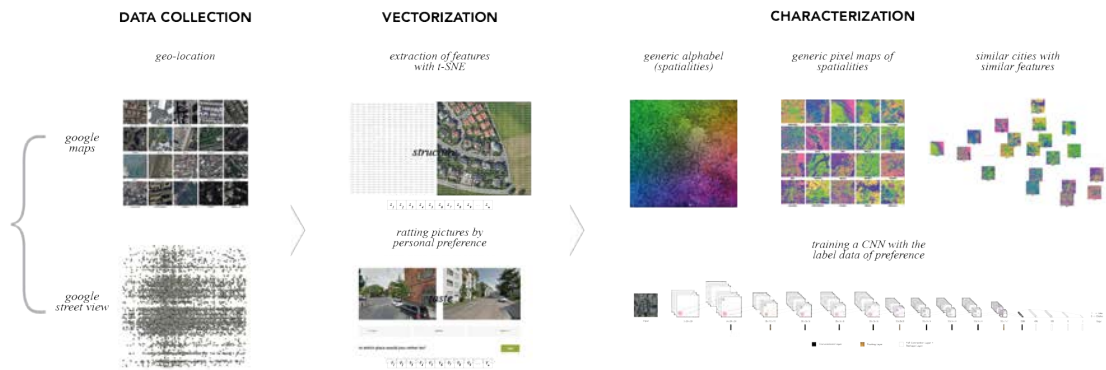

Figure 15. A roadmap explaining the pipeline of the experiment is displayed

The interplay of Machine and Human Intelligence was presented in the chapters The Application: Suggesting new Places to visit and Specific Prediction of Urban Patterns Based on Preference, where these two narrations of the city were coupled as contextual numbers, clustered and characterized once again to generate a personal alphabet of spatialities (fig 16.). These characters were projected back on space using their geolocation to generate Odysseus' Cities of Indexes. These preferences did not reflect a generic understanding of urban likeability but rather the articulation of a personal characterization of the city. With this in mind, a validation of these results is only within the scope of the observer, therefore a generic verification is not desired nor possible.

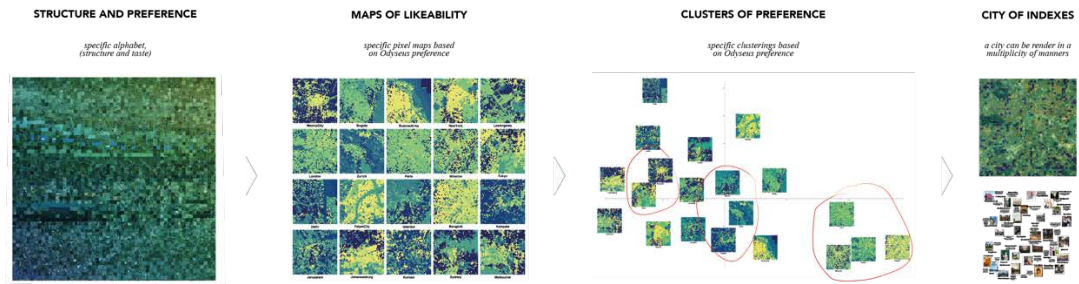

Fig 16. Specific Stage: junction of preference and structure into a personal characterization of the cities or "Cities of Indexes".

In the following images Odysseus' pattern of preference are depicted as a way to confirm its overall consistency. Even if cities are different and unique at the same time, Odysseus may be able to find a similar experience of his preference in a multiplicity of places from different urban cultures. He might like cities over extended areas, like in Johannesburg, Tokyo, Melbourne and Mexico City, with radical transitions across a boundary that may be given by the presence of large natural elements such as the sea or a specific topography. On the other hand, he may think of other cities as melting progressively into their surrounding landscape, likeable all over with some punctual dislikeable places, as it can be seen in cities like Zürich and Istanbul (fig. 17).

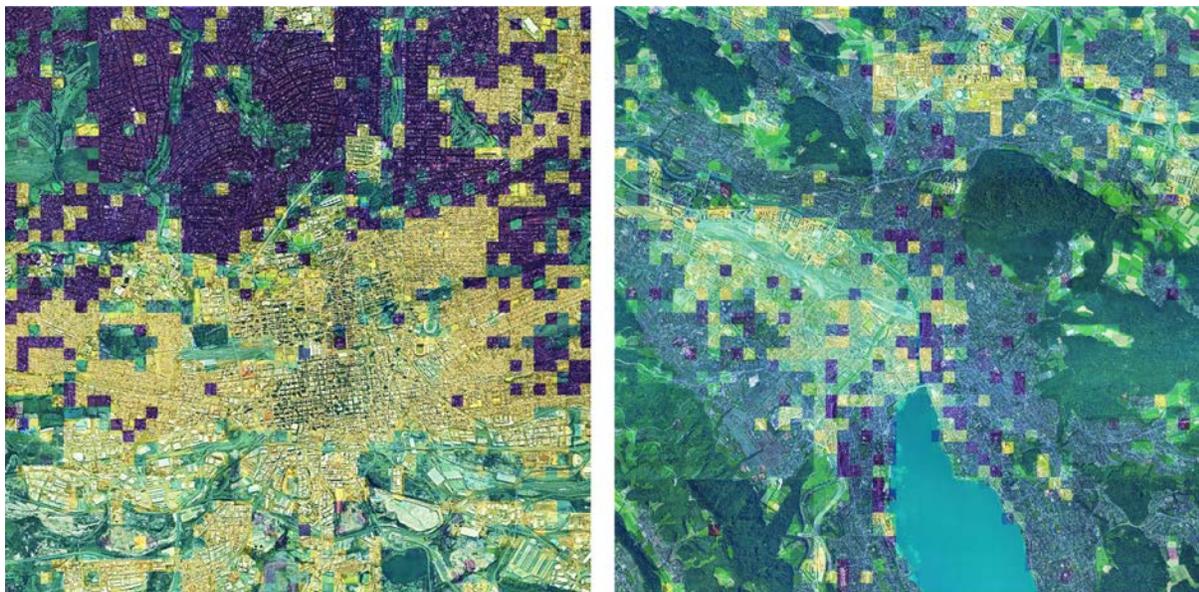

Figure. 17. Johannesburg and Zürich, representing their respective clusters with concentrated patterns of preference.

Inversely, in other cities, namely the ones with an imperial past such as London, Paris and Moscow, he may dislike nearly everything except punctually distributed meaningful places that might refer to monuments or historical landmarks. Other cities, with a higher degree of disorder, such as Kumasi and Delhi, may be likeable as much as dislikeable in a more or less similarly well distributed manner (fig. 18).

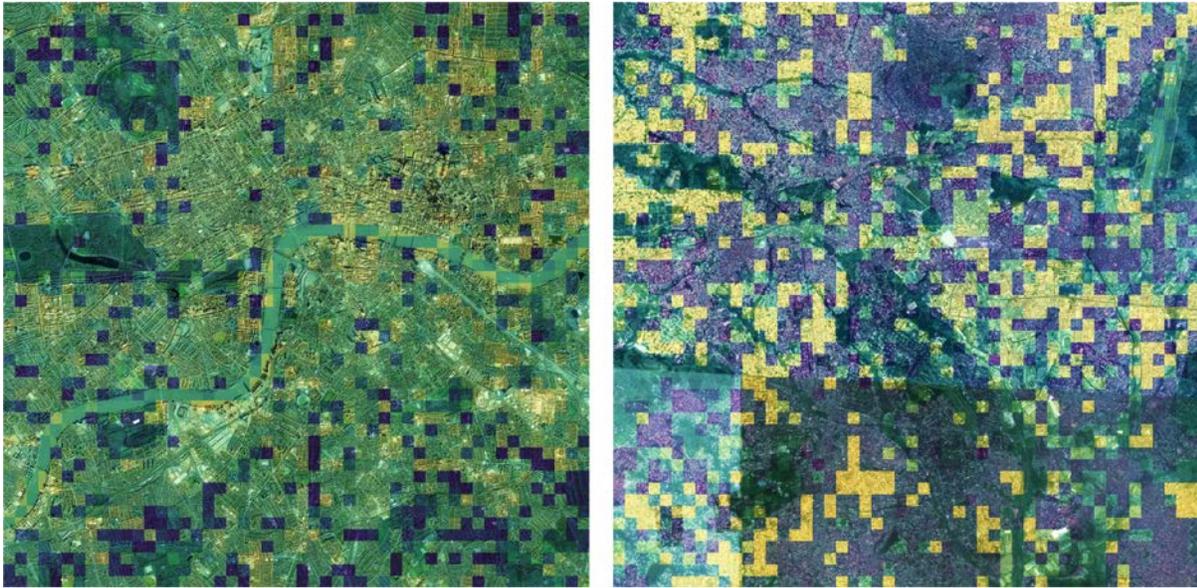

Figure. 18. London and Kumasi, representing their respective clusters with scattered patterns of preference.

Most of the places liked by Odysseus correspond to parks, churches, monuments and cemeteries. Just to name a few, in Johannesburg, the Joubert Park, Johannesburg Art Gallery, the Enoch Sontoga Memorial Park, the Kingsway Campus Auckland Park and the Brixton Cemetery. In Zürich, the Saffa-Insel, the Landiwiese Park, the Mythenquai and the Unterstrass Kirche. In London, the Temple Gardens, the Battle of Britain Memorial, Regent's Park, Battlesea Park and the New Cross Gate area. In Kumasi, the Manhyia Palace, Tafo Asieye (Tafo Cemetery) and Kumasi Children's Park. These Indexical Cities can be rendered in a multiplicity of ways: their location in space is the junction point around which indexes circulate. For the sake of exemplifying, a fictional travelogue of Odysseus' predicted preferences is depicted, at one moment in time of social media's infinite circulation (fig. 19, 20, 21, 22).

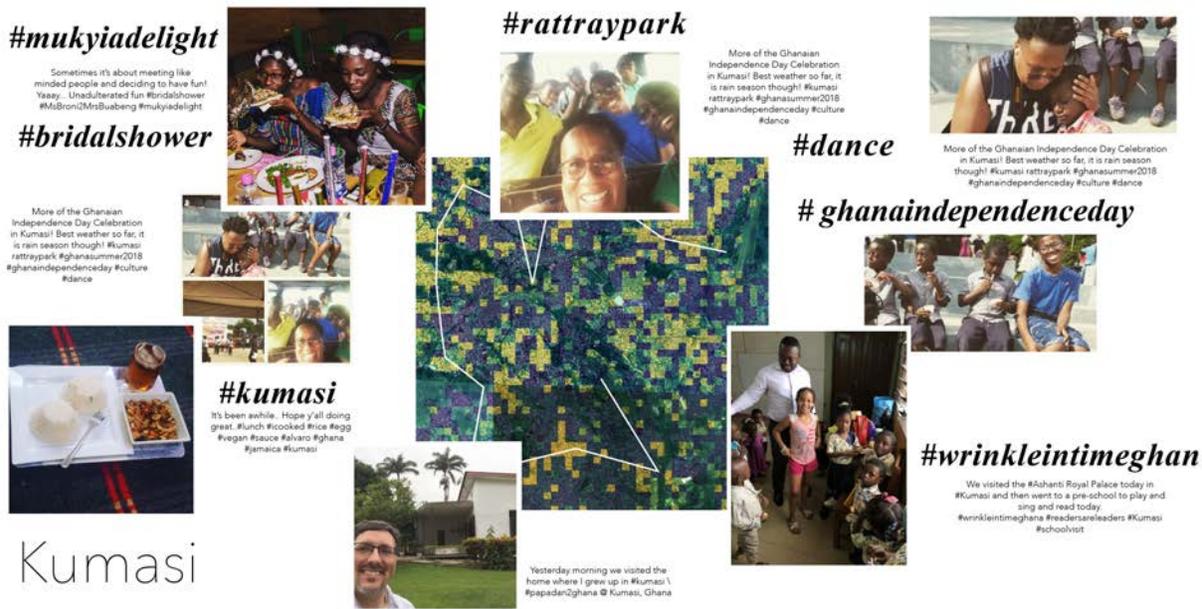

Fig 19. Odysseus' travelogue in Kumasi.

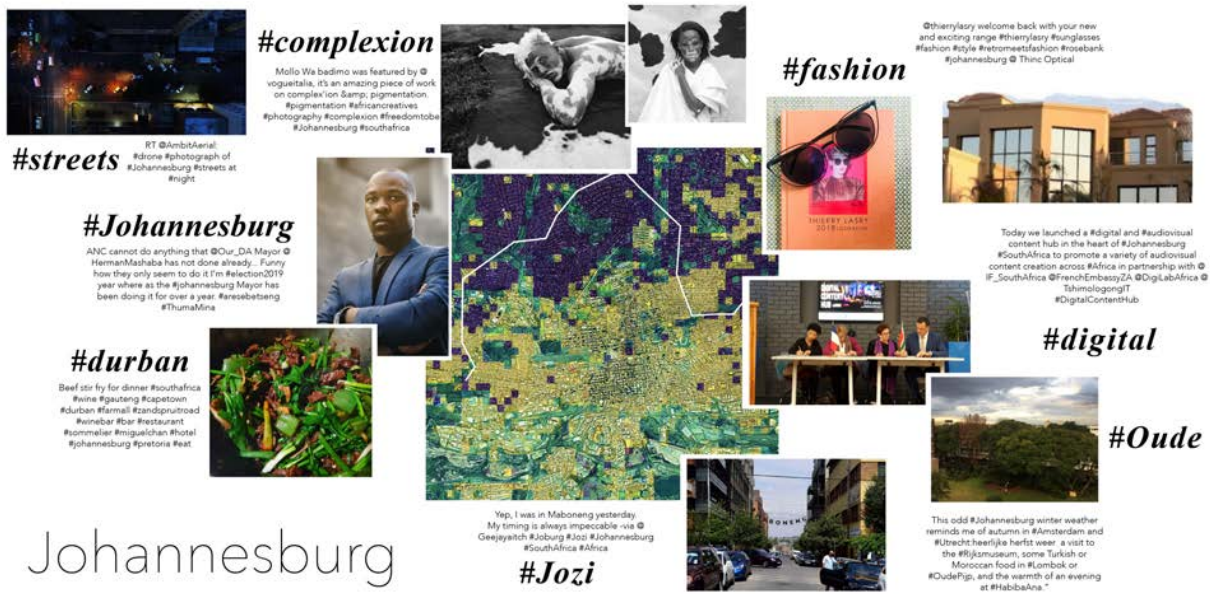

Fig 20. Odysseus' travelogue in Johannesburg.

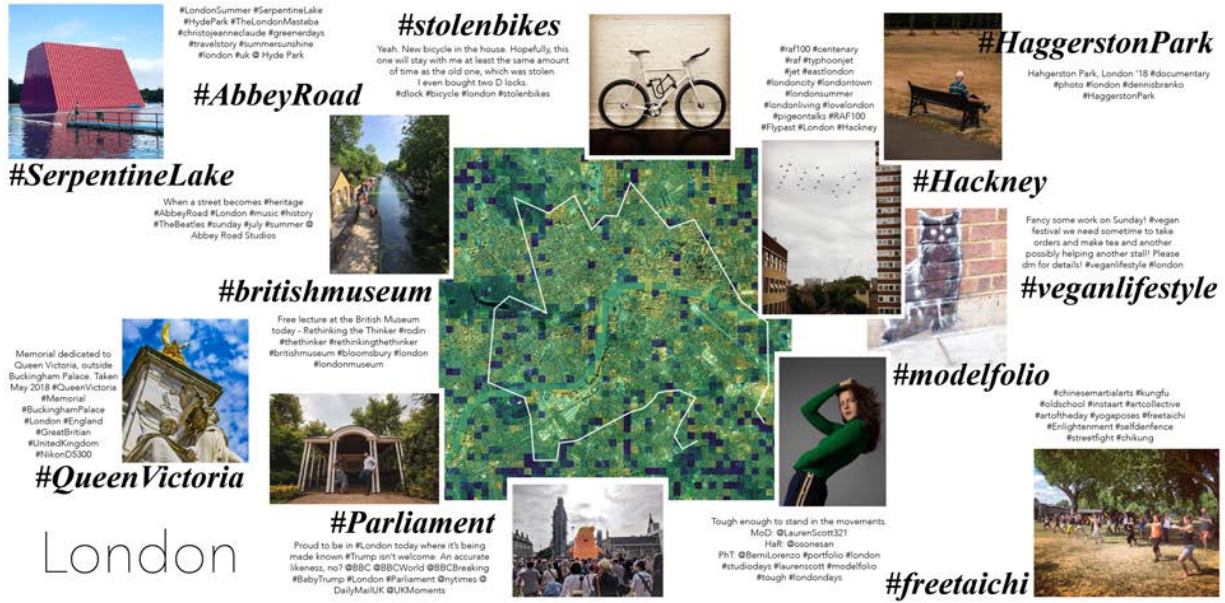

Fig 21. Odysseus' travelogue in London.

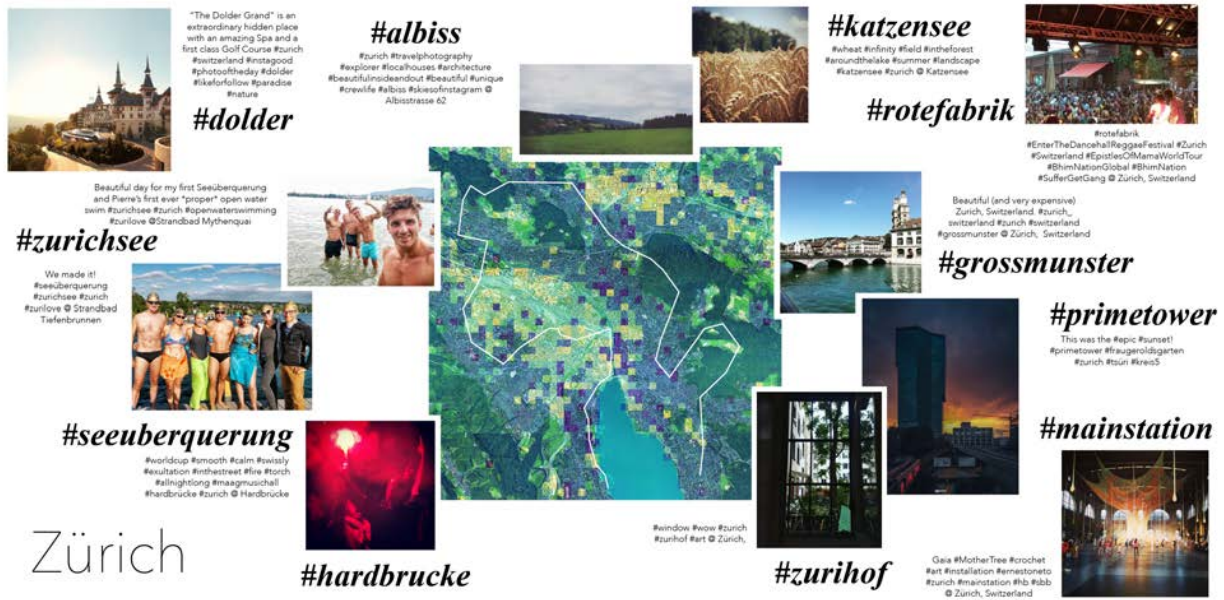

Fig 22. Odysseus' travelogue in Zürich.

However, the purpose of this research was not the specification of such singularities but the constitution of a larger characterization, in the same sense as a theatre character can be thought of. This means, that a character can be thought of as a style or an attitude, in this case, one that depicts how urban places are liked. A characterization, being more abstract than a singular place, remains invariant for one person at a certain given moment in time, yet the manners in which it manifests in cities, are singular, multiple and unique.

6. CONCLUSION

During this experiment, the impossibility of physically perceiving and assessing an unknown place was addressed by predicting the likability of any place in any city on Earth. This obstacle was dealt with by considering such places in an indexical manner, encapsulating different possible representations of places such satellite images and street views simultaneously. Each one of these representations could be symbolized as numerical vectors to be computed with dimensionality reduction clustering algorithms. This research developed two kinds of spatial alphabets: a generic one, corresponding to the clustering of images based on spatial structure (Feature Extraction of an image and its corresponding contextual numbers) and a specific one, corresponding to the personal projection of preference on these images (Classifier output as probability distribution and its corresponding contextual numbers).

With the current abundance of information, the crucial question is neither the axiomatic resolution of problems by establishing a new data science where all responsibility is assigned to Machine Intelligence, nor the upcoming of a new empiricism based solely on Human Intelligence. Machine Intelligence is fast and computationally powerful, and grows exponentially in processing capacity[10], storage capacity[11] and communication capacity1[12]. However, Human Intelligence, deprived of machinic computational power, is capable of articulating complex problems, forming concepts, reasoning, and has the responsibility of taking ethical stances. Each of these forms of intelligence has specific assets and capabilities not to be placed in competition but rather in conjugation.

This experiment started with a generic set of points (50,000 points) and ended with specific sets of points (50,000 characterized points), started with generic pixel maps and ended with specific pixel maps, started with a generic set of images (50,000 satellite images and 32,000 street view images) and ended with specific sets of images (50,000 characterized satellite images and 32,000 characterized street view images).

At the current stage of this research, the resulting application should be seen as a prototype that was only trained with the preferences of the authors, under the fictional character Odysseus. Future steps in this research would direct towards opening up this pipeline to any user who would like train a personal model and identify patterns of personal preferences in urban spaces.

**DATA**
Currently, google satellite images and street views can be download with a developer API. In addition, the datasets, graphics and codes specifically used in this experiment are freely available upon request.

ENDNOTES

[1] Koolhaas, R., Mau, B., 1995. Generic City. S, M, L, Xl.

[2] A convolutional neural network (CNN, or ConvNet) is a class of deep, feed-forward artificial neural networks. CNNs use a variation of multilayer perceptrons designed to require minimal preprocessing. They are also known as shift invariant or space invariant artificial neural networks (SIANN), based on their shared-weights architecture and translation invariance characteristics.

[3] A typical convolution neural network usually consists of two parts: A feature extraction backbone and a classification (or regression) head. The feature extraction part usually consists of many convolution and pooling layers, while the classification head consists of fully-connected layers.

The feature extraction part of the VGG network is an excellent example of transfer learning[4], as it was trained with the broad set of images (ILSVRC2012, 2012) (about 10 million images and 10,000 classes) used for the ILSVRC-2012 challenge.

[5] SOM is a type of artificial neural network (ANN) that is trained using unsupervised learning to produce a low-dimensional (typically two-dimensional), discretized representation of the input space of the training samples.

[6] Steward et al. (2005) postulate that perception can be achieved through the relative judgment (comparison) found in the interdependent dimension among the samples.

[7] Domain adaptation is a field associated with machine learning and transfer learning. This scenario arises when the aim is to learn from a source data distribution a well performing model on a different (but related) target data distribution. (Bridle and Cox 1991) (Ben-David et al 2009)

[8] The accuracy in the recall was 87%, and in precision 90%. The F1Score of the model demonstrated to be consistent.

[9] The BMUs obtained in a linear are known as contextual numbers (Moosavi 2014) and can encapsulate high dimensional data into a continuous one-dimensional numerical field[9], creating an informational face or cypher for each satellite image.

[10] Moore's law the doubling of the data recording capacity on a silicon chip every 18 months. Or every 24 months, depending on the version consulted. Moore's law logically comes up against the physical limits of miniaturization

[11] Kryder's law, which in 2005 predicted that magnetic disk storage density would double every 13 months.

[12] Nielsen's law, connection speed doubles every 21 months.